\documentclass[12pt,preprint]{aastex}

\shorttitle{Criteria for Steady Line-Driven Winds}

\shortauthors{Pereyra}

\begin{document}

\title{Further Criteria for the Existence of Steady Line-Driven Winds}

\author{Nicolas A. Pereyra\altaffilmark{1}}
\affil{University of Pittsburgh, Department of Physics and Astronomy,
       3941 O'Hara ST, Pittsburgh, PA 15260}
\email{pereyra@bruno.phyast.pitt.edu}

\altaffiltext{1}{Universidad Mar\'\i tima del Caribe,
                 Departamento de Ciencias B\'asicas,
                 Catia la Mar, Venezuela}

\begin{abstract}

In Paper~I we showed that steady line-driven disk wind solutions can
exist by using ``simple'' models that mimic the disk environment.
Here I extend the concepts introduced in Paper~I and discuss many
details of the analysis of the steady/unsteady nature of 1D line-driven
winds.
This work confirms the results and conclusions of Paper~I,
and is thus consistent with the steady nature of the 1D
streamline line-driven disk wind models of Murray and collaborators
and the 2.5D line-driven disk wind models of Pereyra and collaborators.
When including gas pressures effects,
as is routinely done in time-dependent numerical models,
I find that the spatial dependence of the nozzle function continues to
play a key role in determining the steady/unsteady nature of supersonic
line-driven wind solutions.
I show here that the existence/nonexistence of local wind solutions
can be proved through the nozzle function without integrating the
equation of motion.
This work sets a detailed framework with which we will analyze,
in a following paper,
more realistic models than the ``simple'' models of Paper~I.

\end{abstract}


\section{Introduction}
\label{sec_introduction}

As discussed in Paper~I
\citep{per04a},
accretion disks are commonly believed to be present in both cataclysmic
variables
(CVs)
and quasi-stellar objects and active galactic nuclei
(QSOs/AGN).
In both types of objects blue-shifted absorption troughs in UV
resonance lines are sometimes present,
giving direct observational evidence for an outflowing wind.
Another property that CVs and QSOs have in common is the existence of a
persistent velocity structure in their absorption troughs
(when present)
over significantly long time scales (Paper~I).
In order for a line-driven disk wind to account for the wide/broad
resonance line absorption structures observed in many CVs and QSOs,
it must be able to account for the steady velocity structure that is
observed.
The 2.5D time-dependent line-driven disk wind models of Pereyra and
collaborators,
both for CVs and QSOs,
have steady disk wind solutions
\citep{per97a,per97b,per00,hil02,per03}.
The earlier 1D line-driven disk wind models of Murray and collaborators
also find steady disk wind solutions
\citep{mur95,mur96,chi96,mur98}.

However,
reports that line-driven disk winds are ``intrinsically unsteady''
(see Paper~I) persist,
based on the argument that the ``unsteadiness'' is physically reasonable
because of the increasing gravity along the streamlines at wind base
that is characteristic of disk winds.
Since the steady nature of CV and QSO wind flows is an observational
constraint,
whether line-driven disk winds are steady or not is a significant issue.
In Paper~I we showed that an increase in gravity at wind base does
{\it not} imply an unsteady wind solution.
We also developed mathematically ``simple'' models that mimic the disk
environment and we showed that
{\it line-driven disk winds can be steady}.

In the past,
evidence in favor of and against steady line-driven disk wind solutions
has come from numerically intensive 2.5D disk wind models.
However,
model differences can be numerical in nature.
Therefore,
a detailed analysis that develops well-defined criteria,
and applies methods independent of the previous numerically intensive
2.5D models,
becomes important.
The goal of this paper is to extend the concepts of Paper~I to establish
such methods and criteria.
In a following paper
\citep[][hereafter Paper III]{per04c},
we will apply the criteria developed here to the exact flux distribution
of a  standard
\citep{sha73}
accretion disk.

I have included gas pressures effects throughout this paper,
rather than assuming gas pressure to be zero as we did in most of
Paper~I for reasons of simplicity.
It has been argued many times in the literature that the assumption of
neglecting gas pressure effects (i.e., zero gas pressure)
in a line-driven wind is reasonable since the main driving force is
radiation pressure.
However,
there are four reasons why it may be important to include gas pressure
effects.

First,
as was discussed by
\citet[][hereafter CAK75]{cas75}
for the stellar case,
and as we discuss in Paper~I and here,
an important property of steady supersonic line-driven winds,
under the Sobolev approximation for the line radiation force,
is that it must have a
``critical point.''
In turn,
the position of the critical point determines the {\it exact}
wind mass loss rates and velocity laws within the model.
The critical point is not just a mathematical construct that
facilitates the calculation of basic wind parameters.
As was shown by
\citet{abb80}
for the stellar case,
the critical point is the point where the flow velocity equals the
backward velocity propagation of density perturbations,
referred to as radiative-acoustic waves or Abbott waves
(rather than sound waves).
Thus,
in a line-driven wind the critical point plays a role physically
equivalent to the sonic point in a temperature-driven wind
\citep{par60}.

However,
if one assumes that gas pressure is zero,
one finds an infinite family of supersonic solutions of which only one
actually has a critical point.
The question that naturally arises from this specific result is:
Does a steady line-driven wind necessarily have to present a critical
point?
The answer is: Yes
(for a wind that reaches supersonic speeds under the Sobolev
 approximation for the line force).
The argument is relatively simple.
As was shown by CAK75 for stellar winds,
no matter how small the gas pressure may be,
the wind solution must have a critical point.
Therefore,
since the zero-gas-pressure case is a limiting case of the
gas-pressure-included case when gas pressure tends to zero,
it follows that only solutions with a critical point are
``real''
or
``physical''
solutions.
Therefore,
the requirement of a critical point is a gas pressure effect.
That is, gas pressure is necessary to identify a unique physical
solution, which further analysis shows it to be a solution that
presents a critical point.

To illustrate the significance of this,
if one assumes that gas pressure is zero
(rather than considering the limiting case when gas pressure tends
 to zero),
the infinite family of possible solutions results in wind mass loss
rates that may vary from any value arbitrarily close to zero up to the
wind mass loss rate of the solution which contains the critical point.
Therefore,
the fact that the wind mass loss rate in a supersonic line-driven wind
arrives at the maximum possible value
(that of the critical point solution)
is a gas pressure effect
\footnote{
In sections 3.1 and 3.2 of Paper~I,
where we neglect gas pressure,
a unique wind mass loss rate is determined because of all the possible
solutions only the critical solution is considered.
In turn,
the reason the critical solution is the only physically acceptable one
is a gas pressure effect,
as can be clearly seen in the Appendix of Paper~I.
}.

This means that although the {\it exact} value of the gas pressure may
not significantly affect the actual value of the wind mass loss rate of
a steady wind,
the existence of gas pressure causes the wind mass loss rate to have
a unique value that corresponds to the critical point type wind
solution.
Thus,
without gas pressure effects,
the wind mass loss rate would not be uniquely determined,
possibly having arbitrarily low values,
which in turn could have important effects on the evolution
of astrophysical systems where line-driven winds are present.

A second motive for considering gas pressure effects is that for cases
where a steady line-driven disk wind solution is not possible
(when assuming gas pressure to be zero),
inclusion of gas pressure effects may allow a uniquely well-determined
solution.
An example of this is the ``S~model'' of Paper I.
If one assumes gas pressure to be zero for the S~model,
then the corresponding nozzle function is monotonically decreasing,
and thus the critical point is at infinity.
Since information,
in principle,
cannot travel an infinite distance in a finite time,
a steady physical solution is not realizable for the zero gas pressure
case.
However,
when gas pressure effects are included,
the corresponding small corrections to the nozzle function produce
a minimum in the nozzle function at a finite distance (Paper~I).
Thus,
once gas pressure effects are considered,
the critical point is no longer at infinity,
but rather at a finite well-defined distance,
allowing for a unique physical solution to be found.
The fact that the inclusion of gas pressure effects may lead a system
from not presenting the existence of a steady solution to presenting
the existence of a steady solution,
is a an obvious and more than sufficient motivation to include gas
pressure effects.

Third,
since evidence in favor and against the existence of steady line-driven
disk wind solutions comes from numerically intensive 2.5D disk wind
models which include gas pressure,
I am unavoidably led to include gas pressure.
If I do not include gas pressure,
for example,
I leave open the possibility that some subtle gas pressure effect at
the wind base could generate strong fluctuations down stream causing
apparent
``intrinsic unsteadiness.''

Fourth,
and finally,
gas pressure is physically present.
For example,
in their line-driven stellar wind paper,
CAK75 did not discuss/justify why they included gas pressure in their
model,
they simply included it since pressure must be present in any
hydrodynamic system.

The specific form of plots,
that would further illustrate the general criteria developed in this
paper,
will depend on the specific models being analyzed.
However,
it is probably best to consider a specific model after the general
framework is discussed.
For this reason,
I show a series of plots corresponding to the CAK75 model,
so as to illustrate and apply the concepts and results of this work
to a well-known well-studied model.
In a following paper (Paper~III),
we shall apply the concepts discussed here to the case of an exact
Shakura-Sunyaev disk flux distribution
\citep{sha73}
and compare the results with the cataclysmic variable disk wind models
of
\citet{per97a}
and
\citet{per00}.

I present a brief discussion of the steady/unsteady vs. the
stable/unstable wind characteristics in
\S\ref{sec_steadyvsstable}.
In \S\ref{sec_equationmotion} I present the 1D hydrodynamic equations
which pertain to the line-driven winds models of this work.
I extend the definition of the nozzle function given in Paper~I to
include temperature gradients in \S\ref{sec_nozzle} and discuss
the critical point conditions in \S\ref{sec_criticalpoint}.
In \S\ref{sec_localsteady} I develop general criteria for the
existence of local steady wind solutions in 1D models and discuss
the existence of global solutions in \S\ref{sec_globalsteady}.
In \S\ref{sec_isocak} I apply the criteria to the well-known and
studied CAK75 stellar wind.
In \S\ref{sec_application} I briefly discuss ongoing efforts and
refer to a following paper
(Paper~III)
in which we will apply the criteria developed here to more realistic
disk wind systems than those previously analyzed in Paper~I.
A summary and conclusions are presented in \S\ref{sec_sumcon}.

\section{Steady Winds vs. Stable Winds}
\label{sec_steadyvsstable}

In general,
a dynamical physical system is said to be steady if and only if
there exists stationary or time-independent solutions to the equations
of motion.
A system is said to be stable if,
in addition to it being steady,
arbitrarily small perturbations to the steady solution will either be
damped or transported  as a wave with constant amplitude by the system.
Thus, systems with steady solutions that amplify arbitrarily small
perturbations are not stable.

Since,
in general,
physical systems are in practice subjected to small perturbations,
the issue of stability is relevant.
In particular,
if a hydrodynamic flow presents a steady solution that is however
unstable, then it will be virtually impossible to find in nature such
a system in its steady state.

\cite{abb80} showed that for the case of 1D line-driven wind,
under the Sobolev approximation for the line force,
small perturbations will travel as radio-acoustic waves
or Abbot waves
(rather than sound waves).
The velocity of the radio-acoustic wave is subsonic in the direction
of the flow and supersonic in the backward direction.
Thus, perturbations to the steady solution of a 1D line-driven will
not be amplified but rather will be transported through the wind,
implying in turn that the wind is stable
\footnote{
In Paper~I,
as an illustrative alternative to the above argument,
and as a consistency check,
we showed that the steady 1D line-driven winds analyzed were stable
through numerical simulations.
Numerical simulations,
due to the finite precision of the calculations,
inherently present small numerical perturbations.
}.

Therefore,
for the specific case of a 1D line-driven wind under the Sobolev
approximation,
if a steady solutions exists,
then it is also stable
\footnote{
Several authors have pointed out that more realistic treatments
of line-force that go beyond the Sobolev approximation may lead to flow
instabilities \citep[e.g.,][]{owo99}.
However,
the Sobolev approximation for the line-force is a standard working
assumption in line-driven wind models,
which is currently being implemented by us and has been implemented
in the numerical models that have reported
``intrinsic unsteadiness''.}.

\section{Equation of Motion}
\label{sec_equationmotion}

Following the notation of Paper~I, 
the 1D hydrodynamic equation of motion for a stationary line-driven
flow is:

\begin{equation}
\left(1 -{b^2 \over 2 W} \right) A {dW \over dz}
=
- \, B A
+ \gamma A \left({A \over \dot{M}} \, { dW \over dz } \right)^\alpha
+ b^2 {dA \over dz}
- A {db^2 \over dz} 
\;\;\;\; ,
\end{equation}

\noindent
which is the equation of motion of Paper~I corrected for temperature
gradient effects;
where
$z$
is the independent spatial coordinate,
$W \equiv V^2/2$
is the kinetic energy per mass,
$V$ is the velocity,
$A$
is the area that depends on
$z$,
$\dot{M} = \rho \, V A$
is the wind mass loss rate,
$\rho$
is the density,
$B$
represents the body forces that corresponds to the gravitational plus
continuum radiation force per mass,
$b$
is the isothermal sound speed,
and
$\gamma$
is the ``line opacity weighted flux''
that also depends on $z$.

As in Paper~I,
I scale the physical parameters by defining a value of
$r_0$,
$B_0$,
$A_0$,
and
$\gamma_0$
as the characteristic distance,
gravitational acceleration,
area,
and line opacity weighted flux,
respectively.
The explicit expressions for the normalized parameters are given in
Paper~I.

Introducing the scaling equations,
the equation of motion becomes

\begin{equation}
\label{equ_motion_w}
  \left(1-{s \over \omega} \right) a {d \omega \over d x}
=
- g a
+ f a \left({a \over \dot{m}} \, {d\omega \over dx} \right)^\alpha
+ 2 s {da \over dx}
- 2 a  {ds \over dx}
\;\;\;\; ,
\end{equation}

\noindent
which is the normalized equation of motion of Paper~I corrected for
temperature gradient effects;
where
$x$
is the normalized independent spatial coordinate,
$\omega$
is the kinetic energy per mass,
$a$
is the area that depends on
$x$,
$\dot{m}$
is the wind mass loss rate,
$g$
is the gravitational plus continuum radiation force per mass,
$s$
is the sound speed squared,
and
$f$
is the ``line opacity weighted flux''
that also depends on $x$.
In equation~(\ref{equ_motion_w}) all physical variables are normalized.

Equivalent to the CAK75 independent spatial variable $u$,
I define the variable $q$,

\begin{equation}
\label{equ_q}
  q
\equiv
  \int\displaylimits_{x_0}^{x} {1 \over a(x')} \, dx' + q_0
\;\;\;\; ,
\end{equation}

\noindent
where $x_0$ is an arbitrary position and $q_0$ an arbitrary value.

\newpage

The equation of motion,
in terms of variables $q$ and $\omega$,
becomes

\begin{equation}
\label{equ_motion_q}
  \left(1-{s \over \omega} \right) {d \omega \over d q}
=
- g a
+ f a \left({1 \over \dot{m}} \, {d\omega \over dq} \right)^\alpha
+ {2 s \over a} \, {da \over dq}
- 2  {ds \over dq}
\;\;\;\; .
\end{equation}

Similar to the CAK75 notation,
I define the function
$h(q)$
as the sum of the independent terms in the equation of motion
(i.e., the terms that are independent of
 $\omega$
 and
 $d\omega/dq$),

\begin{equation}
\label{equ_h}
  h(q)
\equiv
- \, g a
+ {2 s \over a} \, {da \over dq}
- 2  {ds \over dq}
\hskip 48pt
\left[
  =
  - \, g a
  + 2 s {da \over dx}
  - 2 a {ds \over dx}
\right]  
\;\;\;\; .
\end{equation}

\noindent
The equation of motion now becomes

\begin{equation}
\label{equ_motion}
  \left(1 -{s \over \omega} \right) {d\omega \over dq}
=
  h(q)
+ f a \left({1 \over \dot{m}} \, { d\omega \over dq } \right)^\alpha
\;\;\;\; .
\end{equation}

The problem of existence of a steady solution is thus reduced to
determining whether a value of
$\dot{m}$
and a normalized function
$\omega(q)$
exists such that it satisfies the boundary conditions and
equation~(\ref{equ_motion}).
A steady solution for the hydrodynamic 1D model exists if and only if
equation~(\ref{equ_motion}) is integrable while
simultaneously satisfying the boundary conditions.
Typically the boundary condition is the position of the sonic point.
That is,
the equation of motion must not only be integrable,
but additionally there must exist a solution to the equation of motion
that presents a sonic flow speed at a predefined sonic position
(boundary condition).

\section{Nozzle Function and Critical Point}
\label{sec_nozzle}

To continue the discussion I recall the definition of the $\beta$
function given in Paper~I,

\begin{equation}
\label{equ_beta}
  \beta(\omega)
\equiv
  1 - {s \over \omega}
\;\;\;\; .
\end{equation}

\newpage

\noindent
I extend the definition of the nozzle function to include gradient
temperature terms,

\begin{eqnarray}
\label{equ_n}
  n(q)
\equiv
&&
  \alpha (1-\alpha)^{(1 - \alpha) / \alpha}
    {(fa)^{1 / \alpha} \over
     (-h)^{(1-\alpha) / \alpha}}
\hskip 48pt
{\rm for}
\hskip 24pt
  h(q)
<
  0
\nonumber \\
&&
\left[
=
  \alpha (1-\alpha)^{(1 - \alpha) / \alpha}
    {(fa)^{1 / \alpha} \over
     (   g a
       - [2 s / a][da / dq]
       + 2  [ds / dq]
     )^{(1-\alpha) / \alpha}
    }
\right]
\\
&&
\left[
=
  \alpha (1-\alpha)^{(1 - \alpha) / \alpha}
    {(fa)^{1 / \alpha} \over
     (   g a
       - 2 s [da / dx]
       + 2 a [ds / dx]
     )^{(1-\alpha) / \alpha}
    }
\right]
\nonumber
\;\;\;\; .
\end{eqnarray}

The question of existence of a steady solution is reduced to the
question of whether or not,
upon integration, one can always determine

\begin{equation}
{d\omega \over dq}
=
{d\omega \over dq}(q,\omega)
\;\;\;\; .
\end{equation}

Viewing
$dw/dq$
as a function of variables
$q$
and
$\omega$
which satisfies equation~(\ref{equ_motion}),
one can divide the
$q$-$w$
plane into five regions depending on whether a solution for
$d\omega/dq$
exists.
I do this below following the notation of CAK75 for early type stars:

\begin{eqnarray}
\label{equ_region_map}
& & \hbox{Region I:} \; \omega < s \;\; {\rm and} \;\;
    h(q) < 0 \; \hbox{: one solution};
    \nonumber
\\
& & \nonumber
\\
& & \hbox{Region II:} \; \omega > s \;\; {\rm and} \;\;
    h(q) < 0 \;\; {\rm and} \;\; \beta(\omega) \, \dot{m} < n(q) \;
    \hbox{: two  solutions};
    \nonumber
\\
& & \nonumber
\\
& & \hbox{Region III:} \; \omega > s \;\; {\rm and} \;\;
    h(q) > 0 \; \hbox{: one solution};
\\
& & \nonumber
\\
& & \hbox{Region IV:} \; \omega > s \;\; {\rm and} \;\;
    h(q) < 0 \;\; {\rm and} \;\; \beta(\omega) \, \dot{m} > n(q) \;
    \hbox{: no solution};
    \nonumber
\\
& & \nonumber
\\
& & \hbox{Region V:} \; \omega < s \;\; {\rm and} \;\;
    h(q) > 0 \;\; \hbox{: no  solution}.
    \nonumber
\end{eqnarray}

As in Paper~I,
I make the following five assumptions with respect to the solution
$\omega(q)$:
$\omega(q)$
increases monotonically,
the wind starts subsonic,
the wind ends supersonic,
the wind extends towards infinity,
and $d\omega / dq$ is continuous.
Following the arguments presented in Paper~I,
and with the additional assumption that asymptotically

\begin{equation}
q \to \int\displaylimits_{x_0}^{\infty}
        {1 \over a} \, dx'+ q_0
\hskip 24pt
[{\rm i.e.}, \  x \to \infty] :
\cases{ a(q) \to x^2(q) \cr
        s(q) \to s_\infty \hskip 12pt [>0] \cr
        g(q) \, a(q) \to {\rm const} }
\;\;\;\; ,
\end{equation}

\noindent
a solution to the 1D equations must then have the following sequence in
the
$q-\omega$
plane as
$q$
increases
($x$
 increases):

\vskip -13pt

\begin{eqnarray}
\label{equ_region_sol}
& & \hbox{Region I: subsonic,} \; h(q) < 0;
    \nonumber
\\
& & \nonumber \\
\label{equ_cpts}
& & \hbox{Region II: supersonic,} \; h(q) < 0 \; \hbox{, lower branch,}
    \; \beta(\omega) \, \dot{m} < n(q);
    \\
& & \nonumber \\
& & \hbox{Region II/Region IV boundary: supersonic,} \;
          h(q) < 0, \; \hbox{critical point,}
    \; \beta(\omega) \, \dot{m} = n(q);
    \nonumber\\
& & \nonumber \\
& & \hbox{Region II: supersonic,} \; h(q) < 0,  \; \hbox{upper branch,}
    \; \beta(\omega) \, \dot{m} < n(q);
    \nonumber \\
& & \nonumber \\
& & \hbox{Region III: supersonic,} \; h(q) > 0 \ .
    \nonumber
\end{eqnarray}

I extend the definition of a critical point type solution discussed in
Paper~I to solutions of the equation of motion that satisfy the above
conditions [eq.~(\ref{equ_cpts})].
Thus,
as in the isothermal wind case (Paper~I),
I have shown for the assumptions given here that a steady solution must
be a critical point type solution of the equation of motion.

\section{Critical Point Conditions}
\label{sec_criticalpoint}

As discussed in \S\ref{sec_nozzle},
if a steady solution exists for a supersonic 1D line-driven wind
[with the five assumptions on the $\omega(q)$ function described in
 \S\ref{sec_nozzle}],
it must present a critical point.
The critical point is in the boundary between Region~II and Region~IV,
that is,
it is supersonic

\begin{equation}
  \omega_c
>
  s
\;\;\;\; .
\end{equation}

\noindent
The sum of the independent terms of the equation of motion
[see eqs.~(\ref{equ_h}) and (\ref{equ_motion})]
is negative,

\begin{equation}
  h(q_c)
<
  0
\;\;\;\; ,
\end{equation}

\noindent
and it satisfies the condition

\begin{equation}
  \beta(\omega_c) \, \dot{m}
=
  n(q_c)
\;\;\;\; ,
\end{equation}

\noindent
where the subscript
$c$
denotes critical point.

I define

\begin{equation}
\omega'
\equiv
{d\omega \over dq}
\;\;\;\; ,
\end{equation}

\noindent
and following the notation of CAK75,
I define

\begin{equation}
\label{equ_critical0}
  F(q,\omega,\omega')
\equiv
\left(1 -{ s \over \omega} \right) \omega'
- h(q)
- f a \left(1 \over \dot{m}\right)^\alpha
  \left(\omega'\right)^\alpha
\;\;\;\; .
\end{equation}

Then, 
the following three conditions must hold at the critical point:

\noindent
1) the equation of motion

\begin{equation}
\label{equ_critical1}
  F
=
  0
\;\;\;\; ;
\end{equation}

\noindent
2) the critical point at the boundary between Regions II and IV
(\S\ref{sec_nozzle})

\begin{equation}
\label{equ_critical2}
  {\partial F \over \partial \omega'}
=
  0
\;\;\;\; ;
\end{equation}

\noindent
and 3) the continuity of
$\omega'$
(velocity gradients)
[``$dF/dq = 0$''
 combined with eq.~(\ref{equ_critical2})]

\begin{equation}
\label{equ_critical3}
{\partial F \over \partial q}
+ \omega' {\partial F \over \partial \omega}
=
0
\;\;\;\; .
\end{equation}

\noindent
One thus finds three equations
[eqs.~(\ref{equ_critical1})-(\ref{equ_critical3})]
with four unknowns,
namely:
$q_c$,
$\omega_c$,
$\omega'_c$,
and the value of
$\dot{m}$.

Therefore,
the critical point cannot be uniquely determined by
$a(q)$,
$f(q)$,
$g(q)$,
and
$s(q)$.
Thus,
when gas pressure effects are included,
the nozzle function
$n(q)$
cannot by itself determine the exact position of the critical point,
contrary to the case where gas pressure effects are neglected
(Paper~I).

The position of the critical point is determined with an additional
model constraint which is normally the sonic point position.
That is,
the equation of motion,
upon integration from the critical point to lower velocities,
must obtain the correct sonic point.

In \S\ref{sec_appendix} I analyze in detail the critical point
conditions and derive explicit expressions for $\omega_c$,
$\omega'_c$ and $\dot{m}$ as functions of the critical point
$q_c$
(eqs.~[\ref{equ_critical18}-\ref{equ_critical20}]).
Additionally,
in \S\ref{sec_appendix},
I also find that in order for
$q_c$
to be a critical point
(i.e.,
 in order for
 $\omega_c$,
 $\omega'_c$,
 and
 $\dot{m}$
 to be determinable),
it must hold that

\begin{equation}
\label{equ_critical21na}
  h(q_c)
<
  0
\hskip 28pt
  {\rm and}
\hskip 28pt
  \left( {1 \over 2 s_c} \, {ds \over dq} \right)^2
+ {\alpha \over 1 - \alpha} \, {1 \over s_c} \,
  {(-h(q_c)) \over n(q_c)} \, {dn \over dq}
>
  0
\;\;\;\; .
\end{equation}

\noindent
For an isothermal wind ($ds/dq=0$) these two conditions for the
existence of a critical point,
reduce to

\begin{equation}
\label{equ_critical22na}
  h(q_c)
<
  0
\hskip 28pt
  {\rm and}
\hskip 28pt
  \left. {dn \over dq}\right|_{q_c}
>
  0
\;\;\;\; .
\end{equation}

Although the nozzle function
$n$
cannot by itself determine a unique
value for the critical point,
it can constrain the location of the critical point,
and in some cases it can be shown 
{\it without additional calculations}
that a steady solution does not exist
(e.g.,
 in an isothermal wind with a monotonically decreasing nozzle function).

The existence of a critical point
(i.e.,
 the determination of values for
 $\omega_c$,
 $\omega'_c$,
 and
 $\dot{m}$
 [eqs.~(\ref{equ_critical18})-(\ref{equ_critical20})]
 such that the critical point conditions hold
 [eqs.~(\ref{equ_critical1})-(\ref{equ_critical3})])
does not imply that the equation of motion
[eq.~(\ref{equ_motion})]
is locally or globally integrable.
That is,
the existence of a point that satisfies the critical point conditions
does not ensure that a local
(in the vicinity of the point)
steady solution exists or that a global
(throughout the spatial range being considered)
steady solution exists.

In the work of CAK75 for line-driven stellar winds,
the existence of a steady solution was proved by finding
a solution for the particular case of stellar winds.
In the following subsection I show that general criteria for the
existence/nonexistence of local solutions for arbitrary geometries can
be established through the nozzle function
{\it without integrating the equation of motion}.

\section{Requirements for the Existence of Local Steady Solutions}
\label{sec_localsteady}

A local solution in the vicinity of the critical point exists if,
upon integration,
the solution maintains itself in Region II
[see eqs.~(\ref{equ_region_map}) and (\ref{equ_region_sol})].
This is equivalent to the condition that at the critical point

\begin{equation}
\label{equ_local1}
\beta(\omega_c + \Delta \omega) \, \dot{m}
<
n(q_c + \Delta q)
\;\;\;\; ,
\end{equation}

\noindent
where
$\Delta q$
is a variation of
$q$
in the vicinity of
$q_c$,
and
$\Delta \omega$
is the corresponding variation of
$\omega$
determined upon the integration of the equation of motion.

Therefore,
a local steady solution exists if

\begin{equation}
\label{equ_local2}
\left. \left( {d^2 \over dq^2}[\beta(\omega) \, \dot{m} -n(q)]
       \right)
\right|_{q_c}
<
  0
\;\;\;\; .
\end{equation}

I define

\begin{equation}
\label{equ_local3}
\beta'
\equiv
{d\beta \over d \omega}
\hbox{\hskip 12pt}
,
\hbox{\hskip 12pt}
\beta''
\equiv
{d^2\beta \over d \omega^2}
\hbox{\hskip 12pt}
,
\hbox{\hskip 12pt}
n'
\equiv
{dn \over dq}
\hbox{\hskip 12pt}
,
\hbox{\hskip 12pt}
n''
\equiv
{d^2n \over dq^2}
\hbox{\hskip 12pt}
,
\hbox{\hskip 12pt}
\omega'
\equiv
{d\omega \over dq}
\hbox{\hskip 12pt}
,
\hbox{\hskip 12pt}
\omega''
\equiv
{d^2\omega \over dq^2}
\;\;\;\; .
\end{equation}

\noindent
From equation~(\ref{equ_local2}),
if the critical point conditions hold for a given
$q$,
then a local solution in the vicinity of that point exists providing

\begin{equation}
\label{equ_local4}
\beta''(\omega) \, \dot{m} \, (\omega')^2
+ \beta'(\omega) \, \dot{m} \, \omega''
- n''(q)
<
  0
\;\;\;\; .
\end{equation}

For a given
$q$,
the variables
$\omega$,
$\omega'$,
and
$\dot{m}$
can be determined through equations~(\ref{equ_critical18}),
(\ref{equ_critical19}),
and (\ref{equ_critical20}),
respectively.
Considering the definition of
$\beta(\omega)$
[eq.~(\ref{equ_beta})],
$\beta'(\omega)$
and
$\beta''(\omega)$
are given by the following equations:

\begin{eqnarray}
\label{equ_local6}
\beta'(\omega)
&=&
  {s \over \omega^2} 
\;\;\;\; ;
\\
& & \nonumber
\\
\label{equ_local7}
\beta''(\omega)
&=&
  - 2 {s \over \omega^3} 
\;\;\;\; .
\end{eqnarray}

\noindent
The parameter
$n''(q)$
can be calculated from the functions that define the specific model,
namely
$a(q)$,
$s(q)$,
$g(q)$,
and
$f(q)$,
through equations~(\ref{equ_h}) and (\ref{equ_n}).
Additionally,
however,
the evaluation of equation~(\ref{equ_local4}) requires the
determination of
$\omega''$.
This can be obtained as follows.
A solution to the equation of motion must,
of course,
be such that the equation of motion
$F=0$
holds
[eq.~(\ref{equ_motion});
 eq.~(\ref{equ_critical1})].
Therefore,

\begin{equation}
\label{equ_local8}
{dF \over dq}
=
0
\;\;\;\; .
\end{equation}

\noindent
That is

\begin{equation}
\label{equ_local9}
{\partial F \over \partial q}
+ \omega' {\partial F \over \partial \omega}
+ \omega'' {\partial F \over \partial \omega'}
=
  0
\;\;\;\; .
\end{equation}

\noindent
Thus, in general

\begin{equation}
\label{equ_local10}
\omega''
=
{ - {\partial F \over \partial q} 
  - \omega' {\partial F \over \partial \omega}
  \over
  {\partial F \over \partial \omega'}
}
\;\;\;\; .
\end{equation}

\noindent
But,
at the critical point,
both the denominator and the numerator of equation~(\ref{equ_local10})
are equal to zero
[eqs.~(\ref{equ_critical2}) and (\ref{equ_critical3}),
 respectively].
Therefore,

\begin{equation}
\label{equ_local11}
\omega''_c
=
\lim_{q \to q_c}
  { - {\partial F \over \partial q} 
    - \omega' {\partial F \over \partial \omega}
    \over
    {\partial F \over \partial \omega'}
  }
\end{equation}

\noindent
or

\begin{equation}
\label{equ_local12}
\omega''_c
= { - {d \over dq} \left[ \partial F \over \partial q \right]
    - {d \over dq} \left[ \omega' {\partial F \over \partial \omega}
      \right]
    \over
      {d \over dq} \left[ \partial F \over \partial \omega' \right]
  }
\;\;\;\; .
\end{equation}

\newpage

\noindent
Thus,
at the critical point

\begin{equation}
\label{equ_local13}
{d \over dq} \left[ \partial F \over \partial q \right]
+ {d \over dq} \left[ \omega' {\partial F \over \partial \omega} \right]
+ \omega'' {d \over dq} \left[ \partial F \over \partial \omega' \right]
=
  0
\;\;\;\; .
\end{equation}

\noindent
Therefore,

\begin{eqnarray}
\label{equ_local14}
 &  & \left[ {\partial^2 F \over \partial q^2}
             + \omega' {\partial^2 F \over \partial \omega \partial q}
             + \omega'' {\partial^2 F \over \partial \omega' \partial q}
      \right]
\nonumber
\\
& & \nonumber
\\
+ & & \left[ \omega'' {\partial F \over \partial \omega}
             + \omega '
               \left( {\partial^2 F \over \partial \omega \partial q}
                      + \omega' {\partial^2 F \over \partial \omega^2}
                      + \omega''
                        {\partial^2 F \over
                           \partial \omega' \partial \omega}
               \right)
      \right]
\\
& & \nonumber
\\
+ &\;& \omega''
         \left[ {\partial^2 F \over \partial \omega' \partial q}
                + \omega' {\partial^2 F \over
                            \partial \omega' \partial \omega}
                + \omega'' {\partial^2 F \over \partial \omega'^2}
         \right]
=
0
\;\;\;\; ,
\nonumber
\end{eqnarray}

\noindent
which is a second order equation with respect to
$\omega''$.
The partial derivates in equation~(\ref{equ_local14}) can be
calculated through equation~(\ref{equ_critical0}).
The values of
$\omega_c$,
$\omega'_c$,
and
$\dot{m}$
can be calculated through
equations~(\ref{equ_critical18})-(\ref{equ_critical20}).

Given 1D line-driven wind models with arbitrary geometries,
gravitational fields,
flux distributions and temperature structures,
within the approach of this work,
the first step in analyzing the existence of steady solutions
is to determine for which spatial points the critical point conditions
as discussed in \S\ref{sec_criticalpoint} hold.

If there are no points where the critical point conditions hold,
then a steady solution does not exist.
On the other hand,
if these conditions hold for some of the points,
then the next step is to determine which subset of points allow a local
solution
[i.e.,
 satisfy eq.~(\ref{equ_local4})].
If there are no points where local solutions of the equation of motion
are possible,
then a steady solution does not exist.

However,
the existence of local solutions does not ensure the existence of a
global solution.
For example,
if a global minimum of the nozzle function exists farther out than
the set of points that allow a local solution
[and assuming
 $h(q) < 0$
 from the set of local solution points out to the global minimum],
then a global solution does not exist;
this is because the condition
$\beta(\omega) \, \dot{m} < n(q)$
for being in Region~II of the
$q-\omega$
plane
(rather than the nonsolution Region~IV)
breaks down before the solution is extended to the nozzle global
minimum point.
At the critical point
$\beta(\omega_c) \, \dot{m} = n(q_c)$;
and
$\beta(\omega)$
is a monotonically increasing function
[eq.~(\ref{equ_beta})].

As indicated above,
the local solutions in this work refer to local solutions in the
vicinity of a critical point.
The requirement of a critical point for the existence of a steady
solution was discussed in \S\ref{sec_nozzle}.

Thus,
the requirements for the existence of local steady solutions are:

\begin{equation}
\label{equ_local15}
  h(q_c)
<
  0
\;\;\;\; ,
\end{equation}

\begin{equation}
  \left( {1 \over 2 s_c} \, {ds \over dq} \right)^2
+ {\alpha \over 1 - \alpha} \, {1 \over s_c} \,
  {(-h(q_c)) \over n(q_c)} \, {dn \over dq}
>
  0
\;\;\;\; ,
\end{equation}

\noindent
and

\begin{equation}
\label{equ_local16}
\beta''(\omega_c) \, \dot{m} \, (\omega'_c)^2
+ \beta'(\omega_c) \, \dot{m} \, \omega_c''
- n''(q_c)
<
  0
\;\;\;\; .
\end{equation}

\section{Requirements for the Existence of a Global Steady Solution}
\label{sec_globalsteady}

The existence of a global solution
(a solution that spans throughout the spatial range of the model)
requires the existence of a critical point
(\S\ref{sec_nozzle}),
which in turn implies that the critical point conditions must hold
at the given point
(\S\ref{sec_criticalpoint}),
and requires,
of course,
the existence of a local solution
(\S\ref{sec_localsteady})
which can be extended from the critical point towards infinity in the
outward direction and towards the sonic point in the inward direction.

Additionally,
it is required that the global solution be such that upon integration
in the inward direction the wind reaches sound speed at the sonic
point.
The exact position of the sonic point is a boundary condition of the
model
(e.g.,
 in the CAK75 stellar wind model the position of the sonic point is
 assumed to be approximately equal to the stellar photospheric radius).

The set of points for which the critical point conditions hold,
and that allow a local solution,
define the range in which the critical point must be if a global
solution exists.
The exact position of the critical point is determined by initially
guessing with a value within this range,
and iteratively adjusting the critical point position
until the correct sonic point is achieved when integrating inward.
This iterative process to determine the exact critical point position
in a 1D line-driven wind was originally applied by CAK75 in the study of
stellar winds.
Although CAK75 did not present an independent proof of the existence of
local solutions as I do here,
in \S\ref{sec_isocak} I show that in the isothermal CAK75 model,
for the stellar parameters used by CAK75,
the points that allow local solutions extend from the photospheric
radius to beyond 100 times the photospheric radius.
In the original CAK75 model the iterative process was done over a range
of a few integers of the photospheric radius.

Thus,
the requirements for the existence of a global steady solution are
(in addition to the requirements of existence of a local steady
 solution)
that a critical point exists such that upon integration of the
equation of motion,
these following two conditions hold:

\begin{equation}
\label{equ_general1}
  \omega(q_s)
=
  s
\;\;\;\; ,
\end{equation}

\noindent
where $q_s$ denotes sonic point,
and for the points in the supersonic regions where $h(q) < 0$
(other than the critical point $q_c$)

\begin{equation}
\label{equ_general2}
\beta(\omega) \, \dot{m}
<
n(q)
\;\;\;\; .
\end{equation}

\section{Application to the Isothermal CAK75 Stellar Wind}
\label{sec_isocak}

It is well known that when a Sobolev treatment is used for the line
radiation force,
a steady wind solution for early-type stars is obtained
\citep[e.g.,][]{owo99}.
In Paper~I we illustrated concepts and notations by applying them to
the well-known well-studied CAK75 stellar wind,
neglecting effects of gas pressure.
Here I apply the concepts and notations that I present in this
second paper,
but now including gas pressure effects under the assumption of an
isothermal wind.

The purpose of this section in not to shed new light on the CAK75 model,
which is already a well-studied model,
but rather to illustrate the new concepts introduced in this paper
by applying them to a problem familiar to the ApJ reader.
However,
a new result is presented,
that of the proof of existence of a local solution by analytical means
rather than by numerical means.
This is significant because the critical point,
by definition,
is in the boundary between a solution region and a non-solution region,
and therefore numerical methods which typically average parameter values
in the vicinity of a point to extrapolate a function will present
difficulties around the critical point and will have to be adapted
(at the critical point) to the specific case of line-driven winds.

In other words,
subtle numerical methods have to be correctly and consistently
introduced geared specifically to the case of line-driven accretion
winds,
in order to integrate the equation of motion from the critical point.
Since the purpose of this series of papers is to analyze the steady
nature of line-driven disk winds in a form independent of previous
numerical efforts,
the fact that I have found an analytical method to determine the
existence of local solutions,
without integrating the equation of motion,
becomes relevant.
In particular,
CAK75 in their paper,
did not prove the existence of local solutions,
but rather reported that they had obtained one through numerical
integration.
Since the difference between our models and those claiming
``intrinsic unsteadiness''
is a numerical one,
how far our analytical conclusions can be taken becomes important.

As in Paper~I,
I introduce the following characteristic scales for the CAK75 model

\begin{equation}
  r_0
=
  R
\hskip 24pt
,
\hskip 24pt
  B_0
=
  {G M \over R^2} \, (1-\Gamma)
\hskip 24pt
,
\hskip 24pt
  A_0
=
  4 \pi R^2
\;\;\;\; ,
\end{equation}

\noindent
and

\begin{equation}
  \gamma_0
=
  {\kappa_e \over c} \,
  {{\cal L} \over 4 \pi R^2} \,
  k
  \left(1 \over \kappa_e V_{th} \right)^\alpha
\;\;\;\; ,
\end{equation}

\noindent
where $R$ is the photospheric radius,
$G$
is the gravitational constant,
$M$
is the stellar mass,
$\kappa_e$
is the Thomson cross section per mass,
$c$
is the speed of light,
${\cal L}$
is the stellar luminosity,
$k$ and $\alpha$ are the CAK75 line force parameters,
$V_{th}$ is the ion thermal velocity,
and
$\Gamma$
is the Eddington ratio given by

\begin{equation}
  \Gamma
=
  {\kappa_e \, {\cal L} \over 4 \pi G M c}
\;\;\;\; .
\end{equation}

\noindent
For the CAK75 model the independent spatial variable is the distance
$r$
to the center of the star,
and thus
$x=r/R$.

Taking $x_0=1$ and $q_0=-1$,
the variable
$q$
[eq.~(\ref{equ_q})]
becomes

\begin{equation}
\label{equ_isocak_q}
q
=
- \, {1 \over x}
\;\;\;\; .
\end{equation}

\noindent
Equation~(\ref{equ_motion}) becomes the equation of motion,
that is

\begin{equation}
\left(1 -{s \over \omega} \right) {d\omega \over dq}
=
h(q)
+ f a  \left({1 \over \dot{m}} \, { d\omega \over dq } \right)^\alpha
\;\;\;\; ,
\end{equation}

\noindent
where the function
$h(q)$
is given by

\begin{equation}
h(q)
=
- ga
- {4 s \over q}
- 2 \, {ds \over dq} 
\;\;\;\; ,
\end{equation}

\noindent
and in turn

\begin{equation}
  ga
=
  1
\hskip 24pt
{\rm and}
\hskip 24pt
  fa
=
  { 1 \over \alpha^\alpha (1-\alpha)^{1-\alpha} }
\;\;\;\; .
\end{equation}

\noindent
Since I am assuming an isothermal wind
($ds/dq = 0$)

\begin{equation}
\label{equ_isocak_h}
h(q)
=
- 1
- {4 s \over q}
\;\;\;\; .
\end{equation}

For the isothermal CAK75 stellar wind I implement a set of parameters
corresponding to an O5f star,
namely:

\begin{equation}
M
=
60 \, M_{\sun}
\hskip 24pt
,
\hskip 24pt
\Gamma
=
0.4
\hskip 24pt
,
\hskip 24pt
R
=
13.8 R_{\sun}
\;\;\;\; ,
\end{equation}

\noindent
and I use the line force parameters also used by CAK75,
namely:

\begin{equation}
k
=
1/30 
\hskip 24pt
{\rm and}
\hskip 24pt
\alpha
=
0.7
\;\;\;\; .
\end{equation}

\begin{figure}
\epsscale{1.0}
\plotone{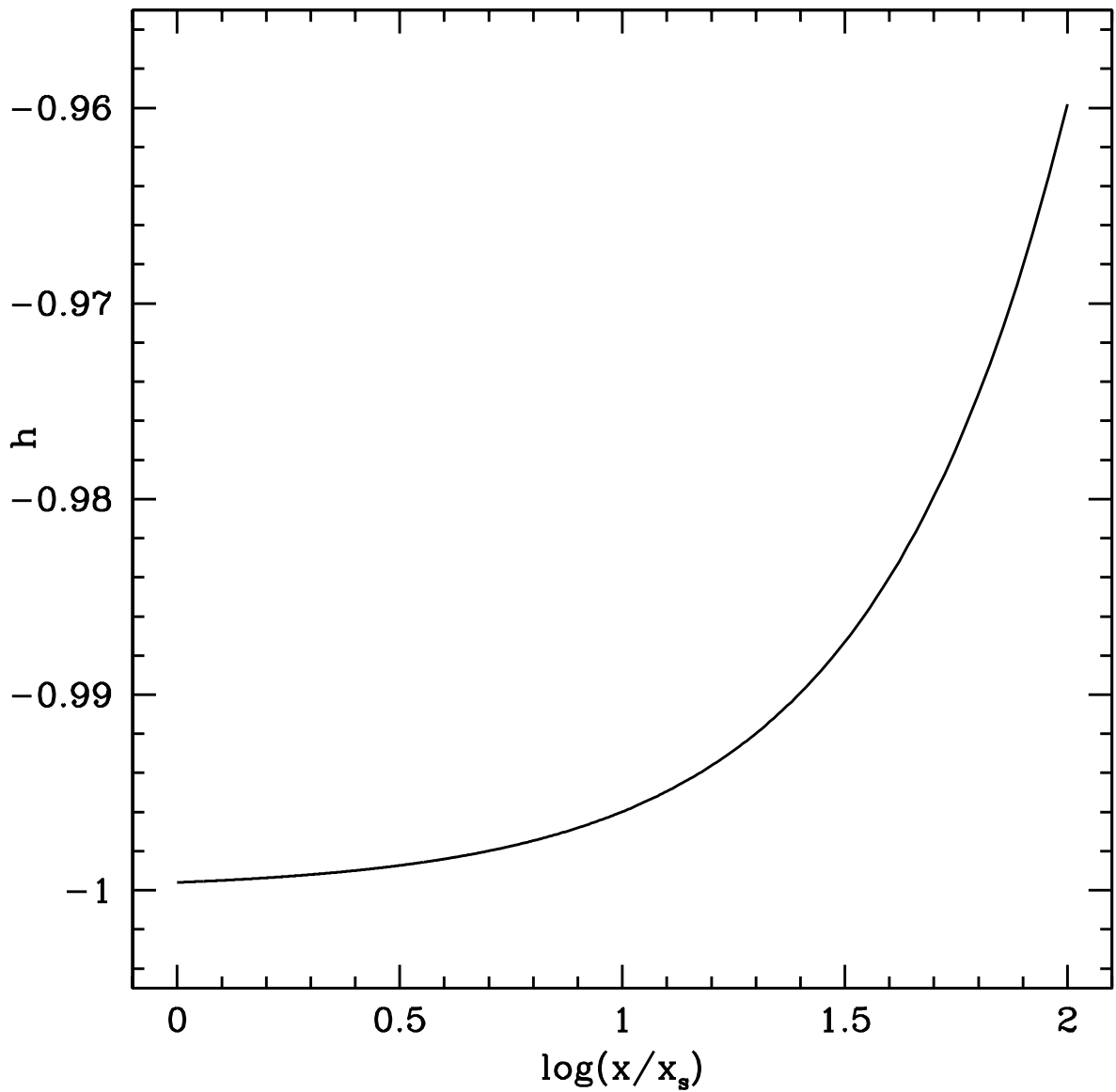}
\caption{
The
$h$
function for the isothermal CAK75 stellar wind.
A necessary requirement for the critical point is that the
$h$
function must be negative at that point.
This figure shows that all points ranging from the sonic point
[$\log(x/x_s) =0$]
to beyond 100 times the sonic height
[$\log(x/x_s) = 2$]
fulfill this condition.
The physical parameters used are:
$M = 60 \, M_\sun$,
$\Gamma = 0.4$,
$R = 13.8R_\sun$
(= photospheric radius),
$k = 1/30$,
and
$\alpha = 0.7$~.
}
\label{fig_cak_h}
\end{figure}

To study the existence of steady solutions,
I first consider the
$h$
function.
As discussed in \S\ref{sec_criticalpoint},
$h$
must be negative at the critical point.
From Figure~\ref{fig_cak_h},
I find that
$h$
is negative from the photospheric height to beyond 100 times the
photospheric radius.
The sonic radius is assumed here to be equal to the photospheric radius.
The maximum possible value for the critical point position,
as constrained by
$h(q)$,
can be determined by equations~(\ref{equ_isocak_q})
and (\ref{equ_isocak_h}).
However,
given the additional assumption that the sonic radius is equal to the
photospheric radius,
I do not expect the critical point to be beyond 100 times the
photospheric radius.

\begin{figure}
\epsscale{1.0}
\plotone{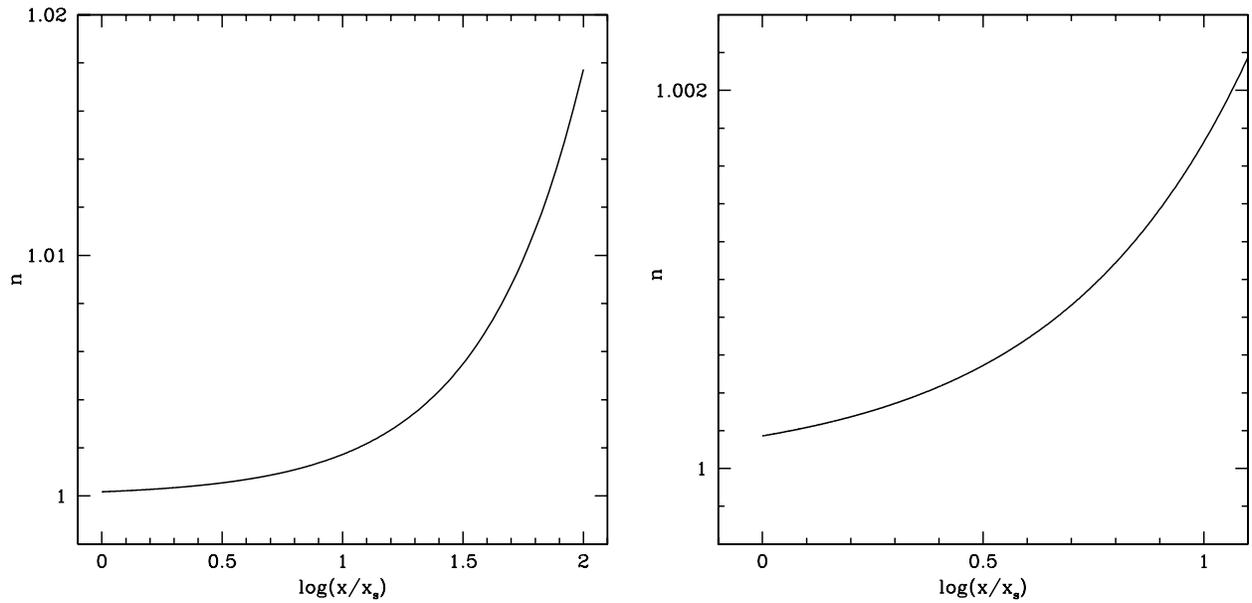}
\caption{
The nozzle function
$n$
for the isothermal CAK75 stellar wind presented at two different scales.
A necessary requirement for the critical point in an isothermal
line-driven wind is that the nozzle function must be locally increasing 
(i.e.,
 $dn/dq > 0$
 [$dn/dx > 0$])
at that point.
This figure shows that all points ranging from the sonic point
[$\log(x/x_s) =0$]
to beyond 100 times the sonic height
[$\log(x/x_s) = 2$]
fulfill this condition.
The physical parameters used are the same as in
Figure~\ref{fig_cak_h}.
}
\label{fig_cak_n}
\end{figure}

Since I am assuming an isothermal wind,
it follows that at the critical point the nozzle function must be
increasing with position
[eq.~(\ref{equ_critical22na})].
I show in Figure~\ref{fig_cak_n} the nozzle function,
and find that the nozzle function is monotonically increasing from
the photospheric height to beyond 100 times the photospheric height.
Thus,
the critical point conditions
(\S\ref{sec_criticalpoint})
hold for all spatial points between the photospheric radius and beyond
100 times this radius.
In other words,
for all these spatial points there are well defined,
well determined values for
$\omega_c$,
$\omega'_c$,
and
$\dot{m}$.

\begin{figure}
\epsscale{1.0}
\plotone{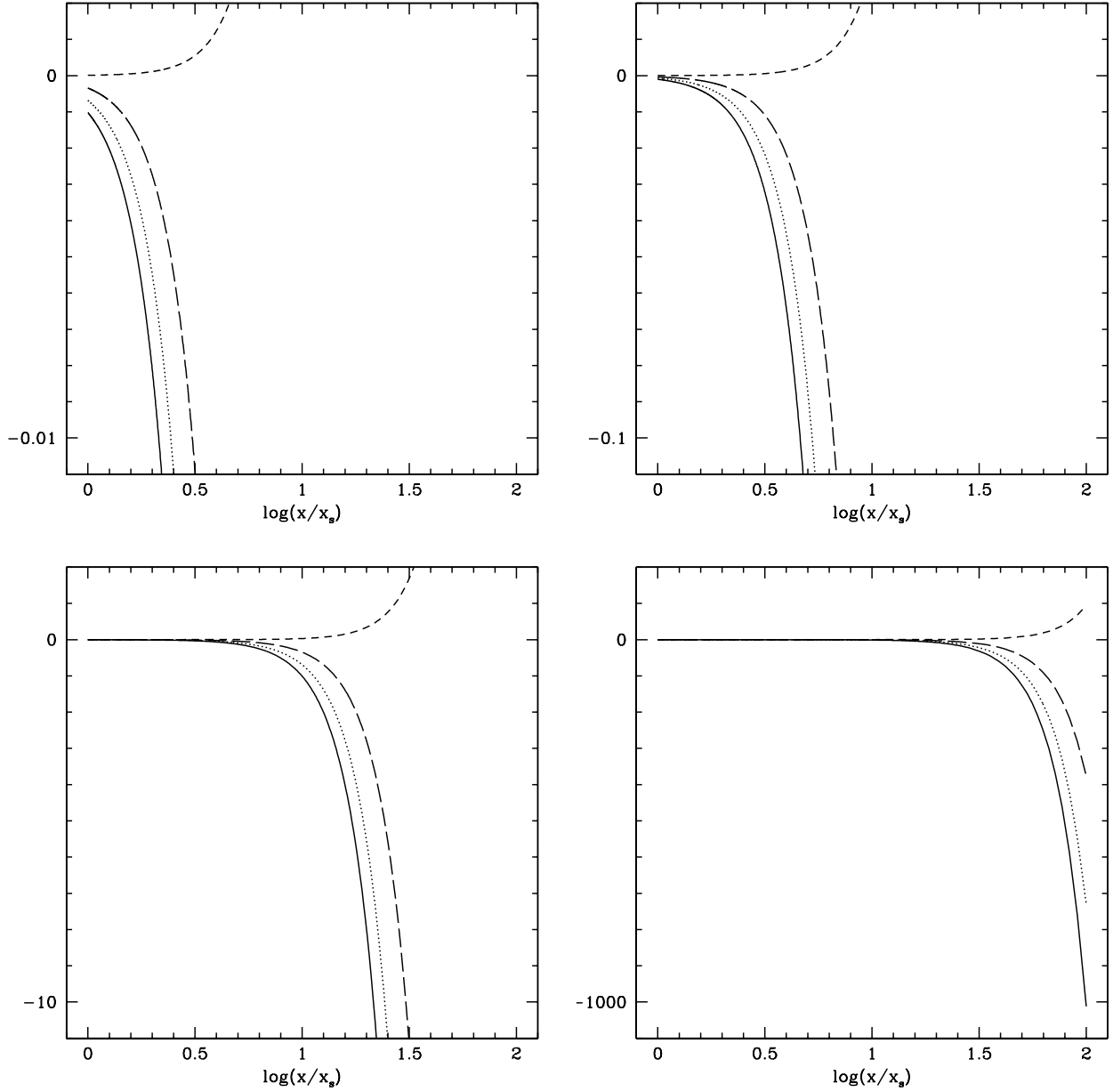}
\caption{
Local solution existence for the isothermal CAK75 stellar wind presented
at four different scales:
a local solution exists in the vicinity of a critical point
$x$
if the expression
$\beta''\, \dot{m} \, (\omega')^2 + \beta' \, \dot{m} \, \omega'' - n''$
(solid curve)
is negative.
This figure shows that all points ranging from the sonic point
[$\log(x/x_s) =0$]
to beyond 100 times the sonic height
[$\log(x/x_s) = 2$]
fulfill this condition.
In this plot the terms
$\beta'' \, \dot{m} \, (\omega')^2$
(dotted curve),
$\beta' \, \dot{m} \, \omega''$
(short dash curve),
and
$-n''$
(long dash curve)
are also shown.
The physical parameters used are the same as in
Figure~\ref{fig_cak_h}.
}
\label{fig_cak_l}
\end{figure}

The existence of local solutions in the vicinity of each critical point 
can be determined through equation~(\ref{equ_local4}).
The expressions
$\beta'$
and
$\beta''$ 
in equation~(\ref{equ_local4}) depend on
the critical point position through the condition
$\omega=\omega_c$~;
and
$\dot{m}$
also depends on the critical point
[eq.(\ref{equ_critical20})].
Figure~\ref{fig_cak_l} shows that all the spatial points up to
100 times the photospheric radius present local solutions in
their respective vicinities.
That is,
throughout the aforementioned spatial range,
the left-hand-side of equation~(\ref{equ_local4}) remains negative.

\begin{figure}
\epsscale{1.0}
\plotone{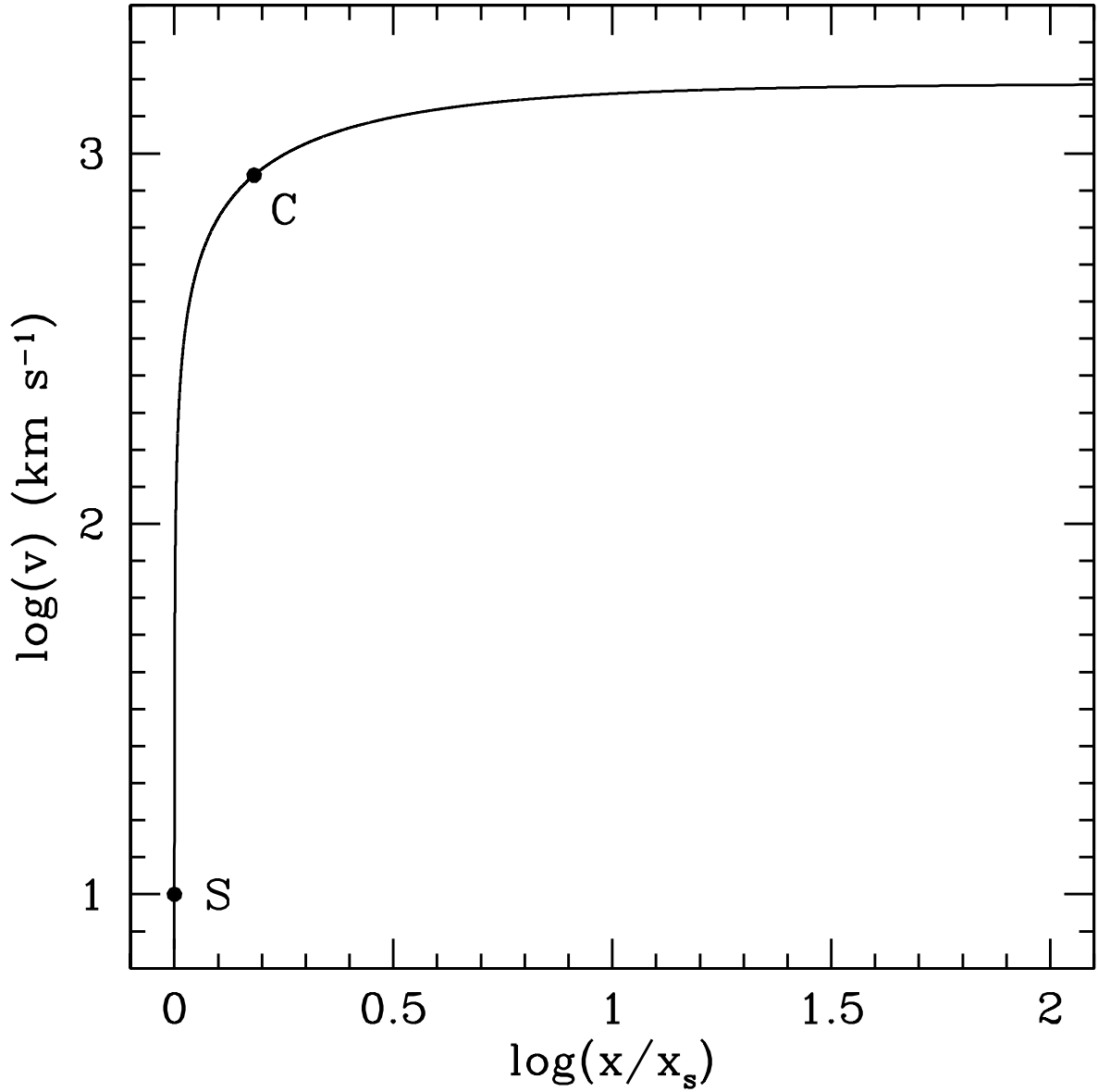}
\caption{
Velocity vs. position
(radius)
for the isothermal CAK75 stellar wind.
The critical point position is determined with the condition that,
upon integration of the equation of motion,
the correct sonic point position is found.
``C''
indicates the critical point and
``S''
indicates the sonic point.
The physical parameters used are the same as in
Figure~\ref{fig_cak_h}.
}
\label{fig_cak_v}
\end{figure}

\begin{figure}
\epsscale{1.0}
\plotone{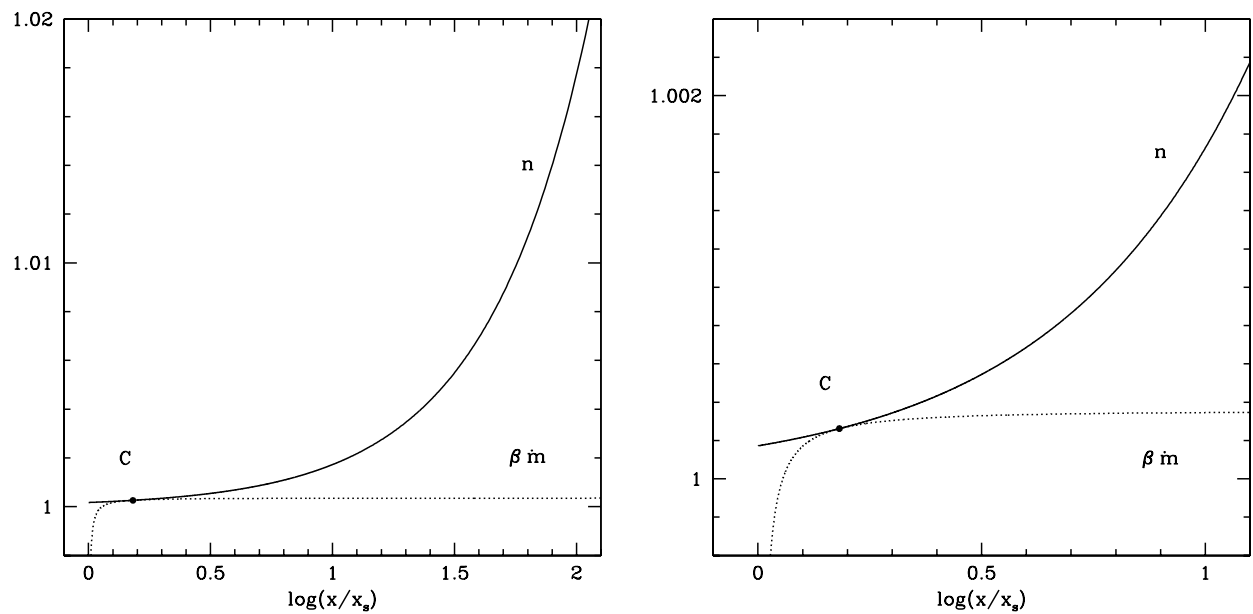}
\caption{
Necessary condition for the global solution existence for the
CAK75 stellar wind for the critical point shown in
Figure~\ref{fig_cak_v}:
upon the integration of the equation of motion,
it must hold that
$\beta(\omega) \, \dot{m} < n(q)$
at points other than the critical point
[when
 $h(q)<0$
 and the wind is supersonic],
and
$\beta(\omega) \, \dot{m} = n(q)$
at the critical point.
Presented here is the nozzle function
$n$
(solid curve)
and the
$\beta \, \dot{m}$
function
(dotted curve)
vs. position at two different scales.
``C''
indicates the critical point.
The physical parameters used are the same as in
Figure~\ref{fig_cak_h}.
}
\label{fig_cak_nb}
\end{figure}

The global solution is found by determining the exact critical
point position by the constraint of the sonic point position as
discussed in \S\ref{sec_globalsteady},
and then integrating the equation of motion inwards and outwards.
Figure~\ref{fig_cak_v} presents the velocity vs. position in the CAK75
stellar wind for the assumed parameters.
As a consistency check for the global solution,
in Figure~\ref{fig_cak_nb},
I plot functions
$n$
and
$\beta \, \dot{m}$.
In the supersonic region while
$h<0$,
for a steady solution to exist,
it must hold that
$\beta \, \dot{m} < n$
for points other than the critical point,
and
$\beta \, \dot{m} = n$
at the critical point.

By comparing the analysis in Paper~1 of a zero-gas-pressure CAK75 wind
with this analysis of an isothermal CAK75 wind,
I find the following two effects of including gas pressure
(in addition to unavoidably complicating the analysis).
First,
in the zero-gas-pressure case the position of the critical point is
undetermined and all points above the photospheric radius become
critical points because the nozzle function is constant;
however,
when gas pressure effects are included,
corrections to the nozzle function arise and a unique critical point is
determined through the conditions discussed in \S\ref{sec_globalsteady}.
Second,
in the zero-gas-pressure CAK75 wind one has to introduce the additional
condition of requiring a critical point in order to uniquely determine
the wind mass loss rate.

Without gas pressure effects there are infinite possible solutions,
each with a different wind mass loss rate.
They vary between the value corresponding to the solution that contains
a critical point,
down to wind mass loss rates arbitrarily close to zero.
When gas pressure effects are included,
only one solution exists,
which is the one that contains the critical point.
Thus,
when gas pressure effects are included,
the physical solution for the equation of motion is found
{\it without}
additional conditions other than the equation of motion itself.

\section{Future Application to Line Driven Disk Winds}
\label{sec_application}

This paper is part of an ongoing group effort to model and analyze the
line driven accretion disk wind scenario for QSOs.
Our first 2.5D model results
[\cite{hil02}]
are encouraging in that they are roughly consistent with QSO
observational constraints.
Among the model/observational agreements is that we are finding
steady/stable winds within our models,
and therefore steady wind line profiles.

However,
reports that line-driven disk winds are ``intrinsically unsteady''
(see Paper~I) persist,
and therefore it becomes relevant to present clear straight forward
evidence,
independent of previous numerically intensive methods,
on whether or not steady/stable line-driven disk wind solutions exist.
If the line-driven disk winds were ``intrinsically unsteady'',
then one would have to discard the line-driven disk wind scenario
on observational grounds.

After considerable efforts and mathematical analysis of the equations,
we have concluded that,
under the Sobolev approximation,
a wind line-driven off a standard accretion disk
\citep{sha73},
is steady.

Given the extensive work that has been done in order to arrive at
this conclusion,
we have decided that it would be best to present it in a systematic
form placing emphasis on the physical ideas on which the results are
based.
In Paper~I,
using simplified models,
we showed that the increase in gravity at wind base along the
streamlines,
which is characteristic of accretion disk winds,
does not imply an unsteady wind.
Further we showed in Paper~I that under reasonable
(but simplified)
conditions {\it line-driven disk winds can be steady}.
The motivation for using the simplified models was to in turn
simplify the mathematical analysis and thus focus on the physical
principles.

However,
Paper~I left one important question unanswered:
if one established an equivalent analysis to more realistic disk
models,
models such as those currently used to numerically simulate 2.5D
line-driven disk winds,
would there continue to exist steady disk wind solutions?
The answer is: Yes.

But,
in order to derive the answer one obviously has to analyze a more
realistic and mathematically more complex system.
Our approach to presenting the results of the more realistic models
has been to first establish a more detailed theoretical/mathematical
framework that would in turn allow us to analyze the more realistic
models.

The objective of this paper is to present this theoretical framework,
and to illustrate it by applying it to the well known CAK75 stellar
wind.
In a forthcoming paper \citep{per04c},
we will apply the framework developed to the flux distribution
corresponding to a standard \citet{sha73} accretion disk.

\section{Summary and Conclusions}
\label{sec_sumcon}

In Paper~I we used ``simple'' models that mimic the disk environment to
show that steady wind solutions can exist.
Paper~I emphasized the underlying physics behind the steady nature of
line-driven disk winds.
The goal of this paper has been to extend the concepts introduced in
Paper~I and discuss important aspects of the analysis of steady/unsteady
1D line-driven winds that were mentioned in Paper~I,
but not discussed in detail.

Specifically,
I show that when including gas pressures effects,
the spatial dependence of the nozzle function continues to play a key
role in determining the steady/unsteady nature of supersonic
line-driven wind solutions.
The existence/nonexistence of local wind solutions can be determined
through the nozzle function without integrating the equation of motion,
as I discuss in detail in \S\ref{sec_localsteady}.
This provides a useful numerical test for models aimed at simulating
line-driven disk winds.

This work sets a detailed framework with which we will analyze more
realistic models than the ``simple'' models of Paper~I.
In a following paper
(Paper~III),
we shall apply the framework discussed here to the case of an exact
Shakura-Sunyaev disk flux distribution and show that if the accretion
disk is capable of sustaining the corresponding wind mass flow,
the wind driven off a standard disk is steady.
This is important in that,
in turn,
it shows that the likely scenario for the formation of absorption
troughs in CVs is a line-driven disk wind.
It also shows that a line-driven accretion disk wind continues to be a
promising scenario to explain the broad absorption lines in QSOs.
In the following paper we shall also compare the results with the
cataclysmic variable disk wind models of
\citet{per97a}
and
\citet{per00}.

\acknowledgments

I wish to thank David A. Turnshek,
Stanley P. Owocki,
Kenneth G. Gayley,
Norman W. Murray,
and 
D. John Hillier
for many useful discussions.
This work is supported by the National Science
Foundation under Grant AST-0071193,
and by the National Aeronautics and Space Administration
under Grant ATP03-0104-0144.

\newpage

\appendix

\section{Analysis of the Critical Point Conditions}
\label{sec_appendix}

The explicit forms of
equations~(\ref{equ_critical1})-(\ref{equ_critical3})
are respectively:

\begin{equation}
\label{equ_critical4}
  \left(1 -{s \over \omega} \right) \omega'
- h
- f a \left(1 \over \dot{m}\right)^\alpha
  \left(\omega'\right)^\alpha
=
  0
\;\;\;\; ,
\end{equation}

\begin{equation}
\label{equ_critical5}
  \left(1 -{s \over \omega} \right)
- \alpha f a \left({1 \over \dot{m}} \right)^\alpha
    \left(\omega'\right)^{\alpha-1}
=
  0
\;\;\;\; ,
\end{equation}

\noindent
and

\begin{equation}
\label{equ_critical6}
  s \left(\omega' \over \omega \right)^2
- {ds \over dq} \left(\omega' \over \omega \right)
- {dh \over dq}
- {d[fa] \over dq}
  \left( 1 \over \dot{m} \right)^\alpha (\omega')^\alpha
=
  0
\;\;\;\; .
\end{equation}

From equations~(\ref{equ_critical4}) and (\ref{equ_critical5}),
respectively,
one then obtains

\begin{eqnarray}
\label{equ_critical7}
  f a \left({1 \over \dot{m}} \right)^\alpha
  \left(\omega'\right)^\alpha
&=& 
  \left(1 -{s \over \omega} \right) \omega'
- h
\;\;\;\; ;
\\
& & \nonumber
\\
\label{equ_critical8}
f a \left({1 \over \dot{m}} \right)^\alpha
  \left(\omega'\right)^\alpha
&=&
  {\omega' \over \alpha} \left(1 - {s \over \omega} \right)
\;\;\;\; .
\end{eqnarray}

\noindent
Therefore,

\begin{equation}
\label{equ_critical9}
  \left(1 -{s \over \omega} \right) \omega'
- h
=
{\omega' \over \alpha} \left(1 -{s \over \omega} \right)
\;\;\;\; ,
\end{equation}

\noindent
which leads to

\begin{equation}
\label{equ_critical10}
{\omega' \over \omega}
=
{\alpha (-h) \over 1 - \alpha} \, {1 \over \omega - s}
\;\;\;\; .
\end{equation}

Equation~(\ref{equ_critical10}) can be rewritten in the form

\begin{equation}
\label{equ_critical11}
  {\omega' \over \alpha} \left(1 - {s \over \omega} \right)
=
  {(-h) \over 1 - \alpha}
\;\;\;\; .
\end{equation}

\newpage

\noindent
Comparing equations~(\ref{equ_critical8}) and (\ref{equ_critical11}),
I have

\begin{equation}
\label{equ_critical12}
  f a \left({1 \over \dot{m}} \right)^\alpha
    \left(\omega'\right)^\alpha
=
  {(-h) \over 1 - \alpha}
\;\;\;\; ,
\end{equation}

\noindent
or

\begin{equation}
\label{equ_critical13}
  \left({1 \over \dot{m}} \right)^\alpha \left(\omega'\right)^\alpha
=
  {1 \over f a} \, {(-h) \over 1 - \alpha}
\;\;\;\; .
\end{equation}

Substituting equation~(\ref{equ_critical13}) into
equation~(\ref{equ_critical6}),
one finds

\begin{equation}
\label{equ_critical14}
{1 \over 2} \left(\omega' \over \omega \right)^2
- {1 \over 2 s} \, {ds \over dq} \left(\omega' \over \omega \right)
- {\alpha \over 1-\alpha} \, {(-h) \over 2 s} \,
  {d \over dq} \left( \ln
                  \left[
                    (fa)^{1 / \alpha} \over (-h)^{(1-\alpha)/\alpha}
                  \right]
               \right)
=
  0
\;\;\;\; .
\end{equation}

Substituting the nozzle function
$n$
[eq.~(\ref{equ_n})]
into the natural logarithm of the third term of
equation~(\ref{equ_critical14}),
one has

\begin{equation}
\label{equ_critical15}
  {1 \over 2} \left(\omega' \over \omega \right)^2
- {1 \over 2 s} \, {ds \over dq} \left(\omega' \over \omega \right)
- {\alpha \over 1 - \alpha} \, {(-h) \over 2 s} \, {d [\ln(n)] \over dq}
=
  0
\;\;\;\; .
\end{equation}

Assuming that the wind is either isothermal or that the temperature
decreases with
$q$
(i.e.,
 decreases with
 $x$),
and taking the positive root of the quadratic
equation~(\ref{equ_critical15}) one finds

\begin{equation}
\label{equ_critical16}
  {\omega' \over \omega}
=
  {1 \over 2 s} \, {ds \over dq}
+ \left[
    \left( {1 \over 2 s} \, {ds \over dq} \right)^2
    + {\alpha \over 1 - \alpha} \, { 1 \over s} \,
        {(-h) \over n} \, {dn \over dq}
  \right]^{1 / 2}
\;\;\;\; .
\end{equation}

Comparing equations~(\ref{equ_critical10}) and (\ref{equ_critical16})
I have

\begin{equation}
\label{equ_critical17}
{\alpha (-h) \over 1 - \alpha} \, {1 \over \omega - s}
=
{1 \over 2 s} \, {ds \over dq}
+ \left[
    \left( {1 \over 2 s} \, {ds \over dq} \right)^2
    + {\alpha \over 1 - \alpha} \, {1 \over s} \,
      {(-h) \over n} \, {dn \over dq}
  \right]^{1 / 2}
\;\;\;\; ,
\end{equation}

\noindent
and thus

\begin{equation}
\label{equ_critical18}
  \omega_c
=
  s
+ {\alpha (-h) \over 1 - \alpha}
  \left\{ {1 \over 2 s} \, {ds \over dq}
          + \left[
              \left( {1 \over 2 s} \, {ds \over dq} \right)^2
              + {\alpha \over 1 - \alpha} \, { 1 \over s} \,
                {(-h) \over n} \, {dn \over dq}
            \right]^{1 / 2}
  \right\}^{-1}
\;\;\;\; .
\end{equation}

\noindent
Substituting equation~(\ref{equ_critical18}) into
equation~(\ref{equ_critical16}),
one finds

\begin{equation}
\label{equ_critical19}
  \omega'_c
=
  s \left\{
      {1 \over 2 s} \, {ds \over dq}
        + \left[
            \left( {1 \over 2 s} \, {ds \over dq} \right)^2
            + {\alpha \over 1 - \alpha} \, {1 \over s} \,
               {(-h) \over n} \, {dn \over dq}
          \right]^{1 / 2}
              \right\}
+ {\alpha (-h) \over 1 - \alpha}
\;\;\;\; .
\end{equation}

\noindent
In turn,
substituting equation~(\ref{equ_critical19}) into
equation~(\ref{equ_critical13}) results in

\begin{equation}
\label{equ_critical20}
  \dot{m}
=
  (fa)^{1 \over \alpha}
  (-h)^{-{ 1 \over \alpha }} (1-\alpha)^{1 \over \alpha}
  \left\{ {1 \over 2} \, {ds \over dq}
           + \left[
               \left( {1 \over 2} \, {ds \over dq} \right)^2
               + {\alpha \over 1 - \alpha} \, s \, {(-h) \over n} \,
                 {dn \over dq}
             \right]^{1 / 2}
           + {\alpha (-h) \over 1 - \alpha}
  \right\}
\end{equation}

\noindent
[cf. eqs. (36),
 (37),
 and (39) of CAK75].

Therefore,
in order for
$q_c$
to be a critical point
(i.e.,
 in order for
 $\omega_c$,
 $\omega'_c$,
 and
 $\dot{m}$
 to be determinable),
it must hold that

\begin{equation}
\label{equ_critical21}
  h(q_c)
<
  0
\hskip 28pt
  {\rm and}
\hskip 28pt
  \left( {1 \over 2 s_c} \, {ds \over dq} \right)^2
+ {\alpha \over 1 - \alpha} \, {1 \over s_c} \,
  {(-h(q_c)) \over n(q_c)} \, {dn \over dq}
>
  0
\;\;\;\; .
\end{equation}

\noindent
For an isothermal wind ($ds/dq=0$) these two conditions for the
existence of a critical point,
reduce to

\begin{equation}
\label{equ_critical22}
  h(q_c)
<
  0
\hskip 28pt
  {\rm and}
\hskip 28pt
  \left. {dn \over dq}\right|_{q_c}
>
  0
\;\;\;\; .
\end{equation}

%\clearpage

\clearpage

\end{document}